\documentclass{ws-procs9x6}            
\usepackage{pdflscape}
\usepackage{color}
\usepackage{colortbl}
\usepackage{booktabs}
\usepackage{graphicx}
\renewcommand{\cite}[1]{\citep{#1}}
\renewcommand{\paragraph}[1]{\noindent\textbf{#1}}
\newcounter{uc}

\definecolor{Grey}{gray}{0.9}
\newcommand{\usecase}[1]{
\refstepcounter{uc} \arabic{uc} &   #1 
}

\begin{document}
\title{Data provenance, curation and quality in metrology}

\author{James Cheney}

\address{University of Edinburgh\\
The Alan Turing Institute\\
$^*$E-mail: jcheney@inf.ed.ac.uk}

\author{Adriane Chapman}

\address{University of Southampton\\
The Alan Turing Institute\\
E-mail: adriane.chapman@soton.ac.uk}

\author{Joy Davidson}
\address{Digital Curation Centre\\
E-mail: joy.davidson@glasgow.ac.uk}

\author{Alistair Forbes}

\address{National Physical Laboratory\\
E-mail: alistair.forbes@npl.co.uk }

\begin{abstract}
Data metrology -- the assessment of the quality of data -- particularly in scientific and industrial settings, has emerged as an important requirement for the UK National Physical Laboratory (NPL) and other national metrology institutes. Data provenance and data curation are key components for emerging understanding of data metrology. However, to date provenance research has had limited visibility to or uptake in metrology. In this work, we summarize a scoping study carried out with NPL staff and industrial participants to understand their current and future needs for provenance, curation and data quality. We then survey provenance technology and standards that are relevant to metrology. We analyse the gaps between requirements and the current state of the art.

\end{abstract}


\bodymatter

%



\section{Introduction}
Metrology, the science  of
measurement, underpins commerce and manufacturing in the UK and
worldwide. International System of Units (SI) provides a coherent foundation for the representation and exchange of measurement data, enabling interoperability and reproducibility in all scientific and technological domains. Concepts such as metrological traceability require a documented, unbroken chain of calibration experiments, each with a stated uncertainty. Confidence in measurement is provided through these traceability chains that establish the provenance of the measurement results.   
As science, engineering and industry increasingly rely
on large-scale data analysis, it is increasingly critical to
adapt the principles of metrology to digital settings  (sometimes called \emph{data metrology}), to ensure confidence in the use of
data.  Provenance, curation, and data quality need to be better
understood in this context.
Over the past 15 years, there has been a great deal of academic
research and development in different scientific communities concerned
with data provenance, data curation, and data quality; however, there seems to be relatively little interaction or mutual awareness between this academic community and the metrology community.

Provenance is information that explains ``where'' some data came from, or ``why'' it was produced in the result of a database query~\cite{DBLP:conf/icdt/BunemanKT01}. Provenance is ``information that helps determine the derivation history of a data product. [It includes] the ancestral data product(s) from which this data product evolved, and the process of transformation of these ancestral data product(s)'' \cite{DBLP:conf/usenix/Muniswamy-ReddyHBS06}. Many overviews of provenance exist, from use in scientific computation and workflow systems \cite{DBLP:conf/ppam/BarkerH07,Bose2005a,Freire:2008,DBLP:journals/sigmod/SimmhanPG05} to types of provenance \cite{cheney2009,Glavic07}, and security over provenance \cite{cheneysecurity14}.  Meanwhile, Digital curation (alias data curation) `involves maintaining, preserving and adding value to digital research data throughout its lifecycle...As well as reducing duplication of effort in research data creation, curation enhances the long-term value of existing data by making it available for further high quality research' \cite{dcc}.  These definitions have been proposed originally with the scientific research data context in mind, but apply equally to data in industry; henceforth we just say `data'.

In 2017--18, we conducted a study initiated by NPL (the UK's National Measurement Institute), interviewing 22 participants from NPL and industry partners in sectors such as energy \& environment, life sciences and health, and advanced manufacturing, all with interests in this area, to  analyze the needs of the metrology community.  This paper summarizes the key findings of this study, and is intended as a guide to existing resources for readers who need \textit{Context and Understanding} of where data came from,  better information for \textit{Curation and Reuse}, \textit{Identify Good Practices},  to perform \textit{Integrity} checks on the data, to provide  \textit{Interoperability}, to facilitate \textit{Linking entities}, to understanding data \textit{Quality}, to enable \textit{Reproducibility} in research, and assess \textit{Uncertainty}. We analyze a large number of use cases drawn from across the metrology domain, and identify their  needs. 

The contributions of this work include:
\begin{enumerate}
    \item A brief survey of provenance and available tools that allow the metrology community to immediately use concepts and tools within their domain, in Section \ref{sec:survey}
     \item Identification of existing data curation and provenance standards that are appropriate for provenance within this domain in Section \ref{sec:standards}.
    \item An analysis of a wide range of use cases from NPL, Advanced Manufacturing, Life Sciences and Health and Energy and Environmental domains, in Section \ref{sec:study}, concluding with suggested next steps.
\end{enumerate}

\section{Overview of provenance and digital curation} \label{sec:survey}
The World Wide Web Consortium (W3C) has developed standards for provenance under the auspices of the Provenance Incubator Group (2010) and Provenance Interchange Working Group (2011--2013).  
 From the W3C Provenance Incubator group, ``Provenance refers to the sources of information, such as entities and processes, involved in producing or delivering an artifact. The provenance of information is crucial in deciding whether information is to be trusted, how it should be integrated with other diverse information sources, and how to give credit to its originators when reusing it'' \cite{provxg}.  Provenance, occasionally referred to as lineage, creates a ``family tree'' of how information, processes and people interacted to create a given artefact\footnote{Artefact, in this context, denotes a digital object such as a data set.}.

Several other related terms are used together with provenance and curation. These include metadata, context and data quality. In this report, we use the following definitions for each term:
\begin{itemize}
\item Metadata is extra information describing a data
  artefact. Based on definitions from the W3C Provenance Incubator Group~\cite{provxg}, ``Descriptive metadata of a resource only becomes part of its provenance when one also specifies its relationship to deriving the resource.'' Thus, metadata information that reflects information about the object, e.g. file size is not provenance information, while metadata that reflects information about creation or modification, e.g. creation date, is provenance information.
\item Context is domain-relevant information about the state of the
  world at the time information is created or modified. Some context
  can be collected with the provenance, e.g., the temperature in the
  laboratory at time of measurement; some can be added later by a user, e.g., annotations; some context is maintained separately,
  e.g., the accuracy of the measurement instrument used.
\item Data quality assessments are estimates of accuracy, precision,
  etc., of the information. These assessments may be automatically or
  manually generated.
\item Annotations can be made on either data or provenance, and are
  typically used to associate context and quality information with
  particular portions of the data and/or provenance.
\end{itemize}

\subsection{Provenance requirements and other considerations}

Groth et al.~\cite{groth12ijdc} identified three distinct dimensions of requirements for provenance information: content, management and use. In this work, we discuss content in Section~\ref{sec:ISprovenance}, while management and use are expanded here along with Capture and Annotation. In addition to these concerns, Allen et al.~\cite{mitreprovenance} outlined practical considerations when designing a system to capture provenance that include capture, annotation, storage, and management. There are core implementation concerns for any system that contains or uses provenance. For instance, a mechanism must exist to gather or \textit{capture} provenance information. This can either be a mechanism, e.g. an API, by which   other systems report provenance information to the provenance  system, or a set of tools that proactively gather provenance   information from other applications.  Additionally, it can be beneficial to associate data quality concerns  and context information with specific portions of provenance  information. Obviously, some thought must be given to how provenance information and annotations are stored for later retrieval and use. Finally, ``Provenance management refers to the mechanisms that make provenance available and accessible in a system" \cite{provxg}, and include managing security and privacy concerns as well as tools for validation, visualisation, representation transformation, etc. \cite{provsuite}.

\subsection{Provenance content \label{sec:ISprovenance}}
Several dimensions related to ``provenance content'' were identified by the W3C community, including: 
\begin{enumerate}
\item Object: The artefact that a provenance statement is about. 
\item Attribution: The sources or entities that contributed to create the artefact in question.  
\item Process: The activities (or steps) that were carried out to generate or access the artefact at hand.  
\item Versioning: Records of changes to an artefact over time and what entities and processes were associated with those changes. 
\item Justification: Documentation recording why and how a particular decision is made.  
\item Entailment: Explanations showing how facts were derived from other facts.  
\end{enumerate}

However, there can be many possible
representations of how processes interacted, or how data was derived, that are equally valid. It is possible to collect large amounts of
provenance data, and still learn nothing from it. So what IS
provenance?  

We know no universal definition, and doubt one exists.  Instead, the 
end usage of the information should be considered when deciding what information should be collected and stored as provenance.   For instance, if an analysis over
the provenance to determine process bottlenecks in a manufacturing
setting is required, then it is imperative that the provenance
contains information about those processes, as well as the starting
and ending timestamps of activities. On the other hand, if the
provenance is to be used to detect anomalous file-access by operating
system processes in a security setting, then the processes and all of
the file-read and file-write actions must be recorded.  Of course, these uses may not always be fully anticipated at the time provenance needs to be recorded, which is why it is important to consider the possible uses and needs at an early stage.

Before any provenance information is recorded, the first consideration
should be ``what is it intended to be used for.'' 
Standardization may be required to identify agreed sets of provenance for particular applications, for example for measuring device metrology this could include uncertainty budget and calibration certificate information.
If the provenance
that is actually captured and stored, cannot be used for the desired
purpose, the systems should be reconsidered. The Use Cases discussed
in Section~\ref{sec:study} involve different sectors/domains and each sector provides
an immediate grounding of what should be in their provenance
system. Additionally, how much provenance is actually required, or the
\emph{granularity} of the provenance, should be carefully
considered. Granularity refers to the level of detail captured. For
instance, in our first example above, the granularity is the set of
processes. In the second example, the granularity is individual
functions (read/writes) within a process. In terms of capture, finer
granularity often requires the capture agents to be embedded deeper in
the applications in order to get the appropriate detail. Additionally,
on the storage side, the higher the granularity, the more provenance
will actually be recorded, and an implementation of the storage component must
be sufficiently scalable to deal with the resulting data size.

\subsection{Digital Curation Lifecycle Model}
\begin{figure}[tb]
  \centering
 \includegraphics[scale=0.7]{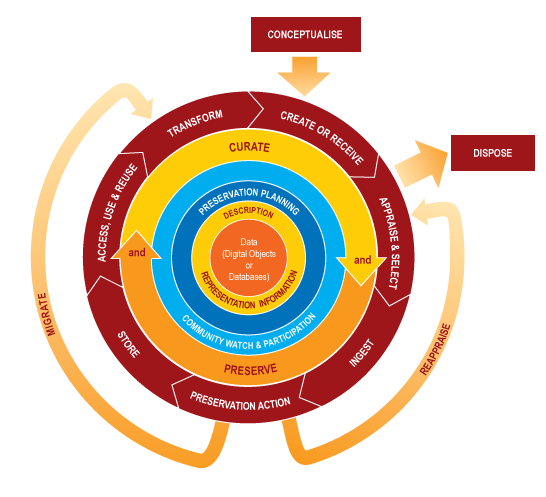}
  \caption{Digital Curation Lifecycle Model \cite{dcclifecycle}}
  \label{fig:curation}
\end{figure}

Digital curation enhances the value of digital information through provision of provenance information, the production of metadata descriptions, and by creating links between outputs to provide context which ultimately supports intelligent reuse. The UK Digital Curation Centre (DCC)'s data-centric Curation Lifecycle Model (Figure~\ref{fig:curation}) was developed to provide a graphical overview of the various stages required for successful curation. We briefly describe of each stage of the digital curation lifecycle; further information can be found on the DCC web site~\cite{dcclifecycle}.

\subsubsection{Digital Curation Lifecycle Components}
During the \textbf{conceptualise} stage, active planning for the creation of outputs should be carried out. Planning should include an assessment of data capture equipment, quality assessment and storage methods as well as any relevant standards that will be adopted. Planning should also consider who needs access to the data during the active stage of data collection and how access will be facilitated.  

During the \textbf{creation stage}, researchers should ensure that sufficient contextual information is captured to make their data FAIR (findable, accessible, interoperable, reusable)\footnote{https://www.force11.org/group/fairgroup/fairprinciples}. Researchers should consider what information may be necessary to validate any published findings. This may include information about hardware used, software or code that is produced and any models or algorithms used to analyse and visualise the data.

Researchers should \textbf{select and appraise} outputs necessary to support reproducibility or for validation purposes but should also be mindful of any additional outputs that may have longer term value. However, not all data generated during a given project can or should be retained.      

Selected outputs should be deposited (\textbf{ingested}) to a suitable digital repository or data centre. Digital Object Identifiers (DOIs) should be assigned to all outputs selected for deposit. The use of identifiers for researchers (e.g.,  ORCiD\footnote{https://orcid.org/}, equipment, funding bodies (e.g., Funder Registry offered by Crossref\footnote{https://www.crossref.org/services/funder-registry/}) and grant numbers should also be used to provide unambiguous context.  

The chosen repository should undertake \textbf{preservation actions} to ensure that data remains authentic, reliable and usable while maintaining its integrity. Researchers may wish to make use of certified repositories (such as CoreTrustSeal\footnote{https://www.coretrustseal.org/}). 

The key objective for digital curation is to support longer term \textbf{access and reuse}. Researchers should consider what context a future reuser may need to understand and make use the given output. FAIR data can be transformed into new research outputs to facilitate new investigations and help researchers to solve grand challenges. 

The derived outputs feed back into the \textbf{conceptualise} stage of the lifecycle model and the cycle begins again. 

\section{Existing resources and standards} \label{sec:standards}

There are many resources, including standards and prototype tools, which may be useful for provenance applications in data metrology.  In this section, we briefly survey these, with references to starting points in the computer science and scientific data management literature.  The coverage we can provide here is necessarily shallow; further details are in the references.

\subsection{Standardization of provenance concepts}
Provenance information exists in many domain-specific standards \cite{iso19115,NMFStandard}. There is also a World Wide Web Consortium (W3C) provenance interoperability standard called W3C PROV~\cite{missier13edbt,w3cprovWeb}. 
The World Wide Web Consortium (W3C) is the main standards body for the World Wide Web, and has created many standards (called Recommendations) including HTML (used for web pages), RDF (a data model for descriptive information using subject-predicate-object triples), and 
OWL, a language for defining ontologies for RDF data.  In this context, ``ontology'' means a set of classes (defining kinds of things), properties (defining kinds of relationships between the things), and rules that impose constraints on the classes and properties, or define ways to infer new knowledge from existing facts.  RDF and OWL are used to represent data in the Semantic Web / Linked Data settings, which have been proposed as ways to make data on the Web more machine-readable and more useful for automated analysis.

The Provenance Interchange Working Group was active from 2011--2013 and produced several W3C Recommendations as well as a number of supporting W3C Notes (documents that support and explain standards, but do not themselves specify standard behaviour). The four main Recommendations are:

\begin{itemize}
\item PROV-DM - which specifies the PROV Data Model, a collection of vocabulary terms used in the PROV standards and their intended meanings\footnote{\url{https://www.w3.org/TR/prov-dm/}}
\item PROV-N - which specifies the PROV Notation (also called PROV-N), a serialization format for PROV information that is intended to be both human and machine readable~\footnote{\url{https://www.w3.org/TR/prov-n/}}
\item PROV-O - which specifies a PROV Ontology (also called PROV-O), an OWL ontology that specifies basic constraints and usage for PROV represented in RDF format~\footnote{\url{https://www.w3.org/TR/prov-o/}}
\item PROV-CONSTRAINTS - which specifies more detailed constraints and inference rules for PROV, and defines a notion of validity and equivalence of PROV datasets~\footnote{\url{https://www.w3.org/TR/prov-constraints/}}
\end{itemize}

Of these standards, PROV-DM and PROV-O are probably the most relevant for typical uses of provenance in the Semantic Web (that is, using RDF and OWL).  Other formats for PROV based on XML and JSON have also been defined, but as supporting documents rather than formal Recommendations.
The design rationale and lessons learned from the development of PROV are described by Moreau et al.\cite{moreau15jws}.  Since PROV has now superseded earlier proposals, we do not go into further details about them.

In the rest of this section, we will summarize the key points of PROV with respect to its potential use in data metrology.

\begin{figure}[t]
  \centering
\includegraphics[scale=0.250]{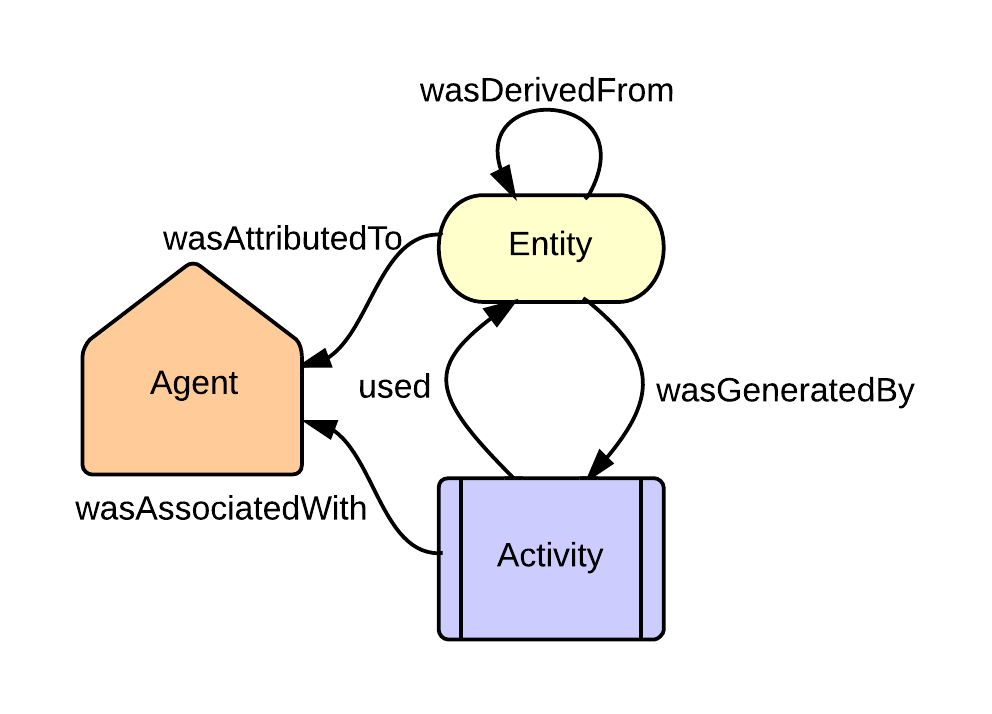}
  \caption{PROV Data Model core concepts and relationships}
  \label{fig:prov}
\end{figure}

Figure~\ref{fig:prov} summarizes the core concepts and relationships of the PROV Data Model.  The three main classes introduced in PROV are:

\begin{itemize}
\item Entities: representing some real-world or informational entity (e.g. a book, an article, a website, an instrument, an image file, a measurement)
\item Activities: representing processes that can be started or ended by agents, and can generate, use, or change entities
\item Agents: representing people, organizations, or software systems that can control or bear responsibility for activities
\end{itemize}

The basic relationships between these concepts are:

\begin{itemize}
\item Usage: an activity may use an entity, meaning its behaviour or outcome is influenced in some way by it
\item Generation: an activity may generate an entity, meaning the entity is created by the activity
\item Association: an agent may be associated with an activity in different ways, such as controlling or participating
\item Information: an activity may inform another activity (typically by some communication)
\item Derivation: an entity may be derived from another entity by an activity.  As a special case, derivation includes versioning where an old version of an entity is replaced by a new one derived from it
\item Delegation: an agent may act on behalf of another agent
\end{itemize}

\begin{figure}[tb]
\centering
\includegraphics[scale=0.4]{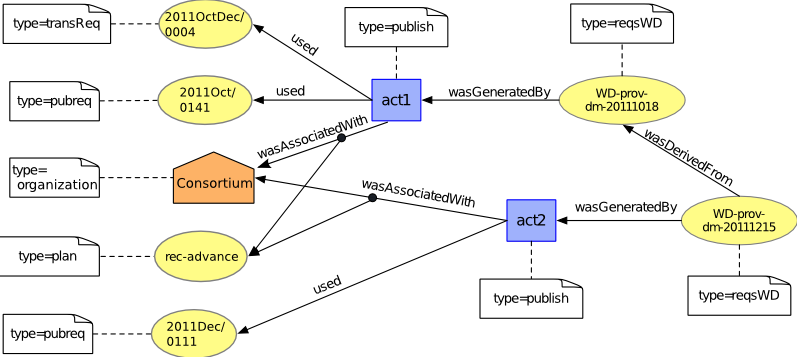}
\caption{Example provenance graph using W3C PROV}~\label{fig:prov-example}
\end{figure}

Using these concepts, it is possible to highlight relationships between artefacts and how they were manipulated. 
A simple example of how the W3C PROV standard might be used to represent provenance of some online documents is shown in Figure~\ref{fig:prov-example}.  

These concepts and relationships are intentionally open-ended and broad.  Typically, a particular application of PROV will make use of additional concepts and relationships suitable to the domain of application.  This is in keeping with the open-ended and distributed nature of the Semantic Web/Linked Data technologies; PROV intentionally does not try to cover all possible application areas, but instead it is encouraged that other vocabularies or ontologies can be used instead (or developed as needed).

For example, to represent executions of scientific computations, the ProvONE vocabulary~\cite{provone} extends PROV with terminology for describing scientific workflow computations.  As another example, to represent the recording and analysis of neuroimaging experiments, the Neuroimaging Data Model (NIDM)~\cite{maumet16scidata} extends PROV with concepts and relationships that standardize terminology relating to neuroimaging instruments and software packages.  

\subsection{Available tools and techniques}
There is a great deal of research on provenance, extending from theoretical aspects \cite{cheney2009,DBLP:conf/icdt/BunemanKT01}, uses \cite{Allen11Collabcom, Chapman11IRI}, tools \cite{provsuite} and implementation concerns \cite{Manish-Anand:2009gb,DBLP:conf/ppam/BarkerH07,DBLP:conf/sigmod/ChapmanJR08}. While much of this research is still in the academic sphere, some has ripened into more mature tools that can be leveraged. In response to the US Federal Government's interest in implementing provenance, and based on the set of questions frequently encountered, the technical report \cite{mitreprovenance} summarizes good practices and engineering design decisions. In the following subsections, we identify many of these more industrial-ready tools. 

\paragraph{Capture tools and technology.}  
There are no generic provenance capture tools that are suited to any kind of application or system. All provenance capture to date involves modifying the underlying system, tools and processes identified as being important to the organisation. In some cases, specific  systems have been modified to provide built-in provenance tracking capabilities; the most relevant example of this approach is scientific workflow management systems, which have been proposed for high-level scientific computation in a number of areas~\cite{DBLP:conf/ipaw/AltintasBJ06,DBLP:conf/sigmod/CallahanFSSSV06,CCPE07-Provenance}. Recently, some of the approaches developed for workflow systems have been adapted to scripting languages such as Python~\cite{yesworkflow,noworkflow}, but this still leaves the problem of how to track provenance when several different systems or languages are used together, each tracking provenance in its own way, or some not at all. Recent work considers a lazy-capture approach, which involves system replay and capture of provenance only when needed \cite{groth2017}.
 
\paragraph{Annotation tools and technology.}  
Annotation allows the enrichment of data by providing additional context or quality information about the underlying data. While it does not necessarily require provenance, it is a piece of metadata that can be attached either to the data, or to the provenance-representation of that underlying data. To date, there are many tools that assist with the creation of annotations and using them. However, most of these tools are very domain specific. For instance, \cite{chemicaltagger} works with chemistry-specific annotations, by highlighting chemistry-specific text, like chemical equations, and providing easy mechanisms for chemistry-specific annotations. Many projects still provide annotation in a do-it-yourself, ad-hoc way. For instance, EMBL-EBI's UniProt\footnote{\url{https://www.ebi.ac.uk/uniprot}} started by custom-building user annotation into their web application, and have since moved on to automatically suggest annotations. More general tools\footnote{Ontotext: https://ontotext.com/} \footnote{Open Calais: http://www.opencalais.com/} focus on ingesting text documents, extracting entities, topic models, social tags, etc., and providing the ability to use and organise this semantic information.

\paragraph{Storage.} 
Provenance is data, and therefore the same storage options exist for provenance as for traditional data including: relational \cite{Manish-Anand:2009gb}; flat file \cite{parkVLDB};  graph-based \cite{missier13edbt}. These storage options are not mutually exclusive. For instance, the data and provenance may be permanently stored in a database, but may be instantiated in a file when passing between systems. Provenance storage introduces a number of challenges, since provenance data can often grow to dwarf the size of the actual data of interest, requiring compression or choices regarding what provenance to keep.  Provenance can also be modelled in different ways, for example as annotations directly on the underlying data, or separately using identifiers for cross-referencing.  The choice of which technology is best for provenance storage will depend upon how that provenance information is to be used, local technology expertise within the organisation, and network and trust architectures within the organisation.  

\paragraph{Administration.}  
Like ordinary data, provenance may need to be secured and protected; in particular, recording provenance in a secure system may introduce new risks to confidentiality (e.g. confidential data might be leaked via provenance) or privacy (e.g. recording user activity may inadvertently record protected personal information). At a minimum, we must apply classic access control techniques to the provenance information \cite{rosenthal09}. However, classic access controls may not be sufficient for provenance, because attackers may be able to leverage knowledge of how provenance is captured to infer missing information even if sanitisation is applied. As such, several protection mechanisms that work more gracefully with provenance have been identified\cite{Hasan2009a,Zhangsdm2009,cheneySecurity,Blaustein11}.  Cheney and Perera~\cite{cheneysecurity14} surveyed and compared a number of techniques in this area.

Beyond security and protection, any provenance solution must have basic administration concerns addressed, including classic considerations for most systems such as facilitating user access, managing users, managing size, scalability, performance, etc.

  \paragraph{Manipulation.} 
   A suite of tools exists to facilitate working with and manipulating W3C PROV~\cite{provsuite}. These include:
 \begin{enumerate}
 \item Validator: A RESTful web service that validates PROV descriptions against the PROV Constraints specification
 \item Translator: Translates between different representations of PROV. Supports PROV-N, PROV-XML, PROV-O and PROV-JSON
 \item Store: A provenance repository that allows storing, browsing, and managing provenance documents via a Web interface or a REST API
 \item ProvToolbox: a Java toolbox for handling PROV
 \item ProvPy: a Python implementation of the PROV data model
 \item ProvExtract: for dealing with PROV embedded in web pages
 \item ProvVis: visualisations of PROV
 \item PROV-N Editor:  a text editor with PROV-N syntax highlighted
 \end{enumerate}

\section{Summary of study: interviews and use cases}\label{sec:study}

As part of our study, individuals from several organisations that
relate to NPL's mission, as well as NPL staff, have been interviewed
in order to identify the scope of possible interest in provenance, and
identify specific use cases and requirements from each area. A brief summary of the interviewed groups and their activities is presented below. These have been organised into the following groups: 
 Activities within NPL; 
 Advanced Manufacturing sector; 
Life Science \& Health sector; 
 Energy \& Environment sector. 

The interviews with NPL staff helped us to understand the current
capabilities and needs within NPL regarding ``data science'' aspects of
data quality, provenance and curation.
For the
interviews with participants from other organisations, or initiatives
where NPL staff are already collaborating with external organisations, we asked the following questions:

\begin{itemize}
\item What is the mission of the organisation/initiative and what role does it
  play in the UK economy?
\item Why are data quality, provenance or curation important to that
  mission?
\item What are the current practices regarding data quality,
  provenance and curation and what are the unmet challenges or future
  needs?
\item What role does NPL currently play, and how could NPL provide
  better support for these needs?
\end{itemize}

For each interview summary, one or more use cases (scenarios
illustrating a challenge or unmet need)  relating to provenance,
curation, and data quality have been identified.  These are listed in Table~\ref{tab:usecases}.  Space limits preclude discussing all of the use cases in detail; instead, we will briefly summarize three representative use cases for each domain.

\paragraph{NPL activities.} We held discussions with several groups within NPL, including staff members representing NPL’s core activities in metrology,such as the measurement of mass, force, pressure, and density, and a staff member in the early stages  of developing a provenance-tracking prototype for stress-strain laboratory data. Based on these discussions we identified several common or significant needs.  Use case 1 involves maintaining records of calibration and measurement methods for measurement equipment, including the measurement procedure, resulting artefacts, calibration methods, and times of (re)calibrations.  Use case 2 considers the possibility that errors might initially go undetected and influence other results, leading to the need to understand how an error has propagated through a system so that users may be alerted and the errors corrected.  Use case 4 considers the reverse scenario, when errors are noticed in derived results and the goal is to diagnose the root cause of the observed error.
All of these use cases raise new challenges as measurements and calibration records move from largely analog, manually-maintained documents recording only a few steps of measurement and computation, to electronic records of possibly long \emph{data supply chains}.

\paragraph{Advanced manufacturing.} 
 Advanced Manufacturing  refers to the sophisticated use of cloud computing, cognitive computing, Internet of Things, and robotics in manufacturing. Because of the reliance on digital models and data analytics, ensuring the quality of the end products also depends upon understanding and extending techniques for data quality, such as tracking provenance through computations, and curation of digital artefacts.  In this sector standards such as the Quality Information Framework (QIF) are used to record manufacturing processes, which is useful for understanding existing processes and where they might break down (Use case 6), and to provide traceability, optimisation, and quality control in manufacturing (Use case 7).  Use case 8 likewise highlights the need to identify and calculate failure points in manufacturing, which may involve working backwards from observations of failed tests to identify root causes, or forward to identify other risks that may be correlated with the root causes identified.

\paragraph{Life science \& health.} 
The Life Science \& Health sector covers medical, pharmaceutical and biological research, as well as healthcare and public health. Use of computation and data management in these areas is already widespread, particularly in biomedical science (‘bioinformatics’). Some NPL researchers are involved in developing computational methodologies that have the potential to improve the quality of patients’ computerised medical records both prospectively by automating the correction of submitted data and retrospectively by reconstructing missing data from historical datasets. These methods enable incomplete or missing data to be made available for analysis and comparison with published results. Use case 9 highlights the potential benefits that AI can offer but introduces issues around transparency as it will be essential that researchers are able to easily distinguish between original data versus those that have been automatically enhanced or reconstructed using algorithms. Many life science researchers depend on having access to reliable physical samples to carry out their analyses. They find that the quality and consistency of the contextual information associated with these physical samples varies greatly which in turn impacts the quality of the analysis they can carry out. Use case 10 highlights that the ability to understand a sample’s quality is directly related to how the physical sample has been sourced and treated prior to its arrival in the lab for digitisation and analysis. NPL researchers are working to improve and standardise sample documentation practices by identifying points over the lifecycle where relevant contextual information is produced and assessing how it should be captured most efficiently and effectively. There is growing interest among researchers in all domains and industry to improve the availability and reuse of existing data. For the life sciences, this may include sharing pre-competitive data across organisations, data collected from hospitals, and subcontracting research to external organisations. Use case 17  introduces a need for improved communication about the context and meaning of the archived data to support informed reuse. As many research performing organisations (RPOs) transition to infrastructure based on cloud storage and object-based data repositories, there is an opportunity to rethink the data capture, curation and management processes to support future reuse.  

\paragraph{Energy \& environment.} 
The Energy \& Environment sector comprises oil \& gas, nuclear, hydroelectric and other
renewable sources of energy, as well as Earth observation, sustainability, climate change impact assessment, and environmental monitoring.  Sensor networks or IoT devices are seeing increasing use for environmental monitoring or monitoring conditions of conventional or renewable energy technologies, but face challenges regarding calibration and uncertainty propagation through data analysis
pipelines. Use case 23 identifies the need for transparency and inter-compatibility of standards used by different groups so that data can be exchanged, compared and reused. Environmental monitoring via Earth observation satellites plays an important role both in understanding the climate and making detailed information about the environment available to government or business, and requires the need to understand and put historical data in context of how it was captured and can be combined with current data, as described in Use Case 21. In a related vein, given the longevity of many environmental studies, it is essential that the provenance survives beyond the lifetime of the software used to create the data or capture the provenance, so that an understanding of this historical data can facilitate data usage later, as discovered in Use Case 22. In all of these cases, the increasing use of large
amounts of data, gathered under varying conditions by devices with varying levels of accuracy, implies the need for provenance and data quality management techniques similar to those that have already been investigated for scientific data management and analysis.

\begin{table}[p]
\tbl{Use cases identified during the study\label{tab:usecases}}{
    \begin{tabular}{|c|p{5cm}|c|c|c|c|c|c|c|c|c|}
 \multicolumn{2}{c|}{}
 & \rotatebox[origin=c]{90}{Context \& Understanding}   
 & \rotatebox[origin=c]{90}{Curation \& Reuse} 
 & \rotatebox[origin=c]{90}{Id Good Practice} 
 & \rotatebox[origin=c]{90}{Integrity} 
 & \rotatebox[origin=c]{90}{Interoperability} 
 & \rotatebox[origin=c]{90}{Linking Entities} 
 & \rotatebox[origin=c]{90}{Quality} 
 & \rotatebox[origin=c]{90}{Reproducibility} 
 & \rotatebox[origin=c]{90}{Uncertainty} \\ 
\hline

\rowcolor{Grey}
\multicolumn{11}{c}{NPL Activities} \\
\hline
\usecase{Understanding Measurements and Calibration Measures \label{use:understandmeasure}}  & $\bullet$ & &$\bullet$ & $\bullet$&$\bullet$ & & & & \\ \hline

\usecase{Taint Propagation \label{use:taint}} & $\bullet$ &$\bullet$   &$\bullet$  &$\bullet$  &$\bullet$  & & & & $\bullet$ \\ \hline
 
\usecase{Analysis of Digital Twins \label{use:digitaltwin}}  & $\bullet$ &   & & & &$\bullet$ & &$\bullet$ & \\ \hline
 
\usecase{Common Cause Analysis \label{use:common}}  &$\bullet$ &   &$\bullet$ & $\bullet$& $\bullet$& & & & $\bullet$\\ \hline

\usecase{Understanding data from stress-strain experiments \label{use:stress}} & $\bullet$ & $\bullet$  &$\bullet$ & &$\bullet$ & & & $\bullet$ & \\ \hline

\rowcolor{Grey}
\multicolumn{11}{c}{Advanced Manufacturing} \\
\hline

\usecase{Modelling and improvement of manufacturing processes \label{use:model}} & $\bullet$ &   & $\bullet$ &  & $\bullet$ & & $\bullet$ & $\bullet$ & \\ \hline

\usecase{Traceability, optimisation, and quality control
in manufacturing \label{use:trace}}  & $\bullet$ &   & & $\bullet$ & & & $\bullet$ &  $\bullet$& \\ \hline

\usecase{Calculation and Identification of Failure Points \label{use:calcfailure}} & $\bullet$ &   & $\bullet$ & $\bullet$ & &$\bullet$ & & & $\bullet$\\ \hline

\rowcolor{Grey}
\multicolumn{11}{c}{Life Science and Health} \\
\hline

\usecase{Automating the reconstruction of missing data \label{use:reconstruct}}  & $\bullet$ &   & & & $\bullet$ & &$\bullet$ & & $\bullet$\\ \hline

\usecase{Linking digital provenance with physical
  artefacts for quality analysis \label{use:linking}}  & $\bullet$&   & & &$\bullet$ &$\bullet$ &$\bullet$ & & \\ \hline

\usecase{Interpretable confidence/quality assessment
  for machine learning \label{use:interpret}}  &$\bullet$ &   & & & & &$\bullet$ &$\bullet$ & $\bullet$\\ \hline

\usecase{Electronic lab notebooks and reproducibility \label{use:notebook}}  & $\bullet$& $\bullet$  & & & & $\bullet$& & $\bullet$& \\ \hline

\usecase{Identifying human gaps \label{use:human}}  & $\bullet$&   &$\bullet$& & & & & & \\ \hline

\usecase{Standards and best practices for reuse \label{use:reuse}} & $\bullet$& $\bullet$  & & & & & & & \\ \hline


\usecase{Facilitating communication and understanding across  organisations \label{use:communication}}  & $\bullet$&   & & & $\bullet$ & & $\bullet$ & & \\ \hline


\usecase{Provenance and Quality in trials \label{use:trials}}  & $\bullet$ &   &$\bullet$ &$\bullet$ & $\bullet$& & $\bullet$& & \\ \hline

\usecase{Data Reuse for additional purposes \label{use:newpurpose}}   & $\bullet$ & $\bullet$  & $\bullet$ & $\bullet$ & & & & & \\ \hline

\usecase{Identifying duplication in rare genetic data \label{use:duplication}}  & $\bullet$& $\bullet$  & & & & $\bullet$ &$\bullet$ & & $\bullet$\\ \hline

\usecase{Diagnosing unexpected differences in
  results of analytical models \label{use:differences}} & $\bullet$ & &$\bullet$ & & &$\bullet$ & $\bullet$ & &$\bullet$ \\ \hline

\usecase{Understanding changes to data over time \label{use:time}}   & $\bullet$ & $\bullet$ & & $\bullet$& & & & & \\ \hline

\rowcolor{Grey}
\multicolumn{11}{c}{Energy and Environment} \\
\hline

\usecase{Reuse and understand historical data with recent data \label{use:history}} & $\bullet$&$\bullet$ & & $\bullet$&$\bullet$& &  & &$\bullet$ \\ \hline

 \usecase{Recording and using past configurations and performance \label{use:container}} & $\bullet$ & $\bullet$ & & & $\bullet$& & & & \\ \hline

\usecase{Transparency and inter-compatibility of standards \label{use:transparency}}  & $\bullet$ &   &$\bullet$ & & & & & & \\ \hline
 
  \usecase{Reflecting good practice in new tools and services } & & & $\bullet$ & $\bullet$ & & & & & \\ \hline

\end{tabular}
}
\end{table}

\subsection{Recommendations}

We made several recommendations to NPL based on our study, which we have adapted for consideration by the broader metrology community as follows:
\begin{itemize}
\item \textbf{Metrologists should establish best practices for provenance standards} to
  provide greater clarity regarding what techniques are sufficient to
  meet current or future needs.
\item \textbf{Flagship case studies should be developed} to consolidate expertise with developing and using provenance techniques.
\item \textbf{Metrology institutes should raise awareness of provenance as part of data metrology}
  through augmenting training materials or other dissemination
  pathways.
\item \textbf{The importance of data management planning should be highlighted} to
  promote data reuse and curation in
  industrial settings.
\item \textbf{Academic research and industry need to be brought together} to facilitate
  transfer of solutions to practice, or communicate unmet needs to researchers.
\end{itemize}
Addressing these recommendations is the subject of current work by NPL and  collaborators within the metrology and data science communities. For example, the EURAMET Technical Committee for Inter-disciplinary Metrology (TC-IM) has initiated a project on Research Data Management and the Consultative Committee for Units (CCU) of International Bureau for Weights and Measures (BIPM) has set up a task group (D-SI) to promote the digitisation of metrology data exchange, references, services and products and to transform the SI into the digital world. In these initiatives, it is seen as essential to address issues relating to data provenance, curation and the FAIR principles and that the metrology community joins with the wider scientific and technical communities to provide well-founded, coherent and inter-operable solutions.

\section{Conclusion}

Our study highlighted that there are a number of situations in metrology and industrial settings where provenance information is needed, and thus there are many settings where techniques developed by the provenance research community could be beneficial, but also showed that there is relatively little awareness or adoption of these techniques.  This may in part be because of the initial focus of the provenance research community on settings such as scientific data and computation and on Web technologies that seem not to be in widespread use in industry settings.  For example, only in the life science and health sector (where RDF is already widely used) did we find pre-existing awareness of RDF and PROV.  Thus there remains a significant gap between what is technically possible in research prototypes and the day-to-day needs expressed in the use cases we encountered.  We hope that this paper will serve to catalyze further discussion between the metrology and data provenance communities to address these challenges.

\paragraph{Acknowledgments}
This work was supported by EPSRC (grant number EP/S028366/1) and an ISCF Metrology  Fellowship grant provided by the UK government’s Department for Business, Energy and Industrial Strategy (BEIS).

{\small
\bibliographystyle{ws-procs9x6} 
\bibliography{provxg,study}
}
\end{document}